 \documentclass[aps, onecolumn, showpacs, amsmath]{revtex4-2}
\usepackage{dcolumn}% Align table columns on decimal point
\usepackage{bm}% bold math
\usepackage{graphicx}% Include figure files
\usepackage{color}
\usepackage{hyperref}
\usepackage{amsthm}
\usepackage{amssymb}
\begin{document}
\def\be{\begin{equation}}
\def\ee{\end{equation}}

\def\bc{\begin{center}}
\def\ec{\end{center}}
\def\bea{\begin{eqnarray}}
\def\eea{\end{eqnarray}}
\newcommand{\avg}[1]{\langle{#1}\rangle}
\newcommand{\Avg}[1]{\left\langle{#1}\right\rangle}

\def\ie{\textit{i.e.}}
\def\etal{\textit{et al.}}
\def\m{\vec{m}}
\def\G{\mathcal{G}}

\newcommand{\davide}[1]{{\bf\color{blue}#1}}
\newcommand{\gin}[1]{{\bf\color{green}#1}}

\title{Extremal statistics for first-passage trajectories of drifted Brownian motion under stochastic resetting}
\author{Wusong Guo}
\author{Hao Yan}
\author{Hanshuang Chen}\email{chenhshf@ahu.edu.cn}
\affiliation{School of Physics and Optoelectronic Engineering, Anhui University, Hefei 230601, China}
\begin{abstract}
We study the extreme value statistics of first-passage trajectories generating from a one-dimensional drifted Brownian motion subject to stochastic resetting to the starting point with a constant rate $r$. Each stochastic trajectory starts from a positive position $x_0$ and terminates whenever the particle hits the origin for the first time. \textcolor{blue}{We obtain the exact expression for the marginal distribution $P_r(M|x_0)$ of the maximum displacement $M$}. We find that stochastic resetting has a profound impact on $P_r(M|x_0)$ and the expected value $\langle M \rangle$ of $M$. Depending on the drift velocity $v$, $\langle M \rangle$ shows three distinct trends of change with $r$. For $v \geq 0$, $\langle M \rangle$ decreases monotonically with $r$, and tends to $2x_0$ as $r \to \infty$. For $v_c<v<0$, $\langle M \rangle$ shows a nonmonotonic dependence on $r$, in which a minimum $\langle M \rangle$ exists for an intermediate level of $r$. For $v\leq v_c$, $\langle M \rangle$ increases monotonically with $r$. Moreover, by deriving the propagator and using path decomposition technique, we obtain in the Laplace domain the joint distribution of $M$ and the time $t_m$ at which the maximum $M$ is reached. Interestingly, the dependence of the expected value $\langle t_m \rangle$ of $t_m$ on $r$ is either monotonic or nonmonotonic, depending on the value of $v$. For $v>v_m$, there is a nonzero resetting rate at which $\langle t_m \rangle$ attains its minimum. Otherwise, $\langle t_m \rangle$ increases monotonically with $r$. We provide an analytical determination of two critical values of $v$, $v_c\approx -1.69415 D/x_0$ and $v_m\approx -1.66102 D/x_0$, where $D$ is the diffusion constant. Finally, numerical simulations are performed to support our theoretical results.    	
\end{abstract}

\maketitle
\section{Introduction}
Despite infrequent occurrences, extreme events have profound impact on almost all
event series in our lives, ranging from natural calamities like earthquake, tsunamis and floods to economic collapses and outbreak of pandemic \cite{fisher1928limiting,gumbel1958statistics,leadbetter2012extremes,bouchaud1997universality,davison2015statistics,albeverio2006extreme}. Extreme value statistics (EVS) has been a branch of statistics which deals with the extreme deviations of a random process from its mean behavior. The EVS of identically distributed random variables is
now well understood, thanks to three distinct universality classes depending on the tails of the distribution of random variables, namely, Gumbel, Fr\'echet, and Weibull  \cite{gnedenko1943distribution}. In recent years, there is an increasing interest in studying the EVS for weakly and strongly correlated stochastic processes \cite{majumdar2010universal,schehr2014exact,lacroix2020universal,PhysRevLett.111.240601,PhysRevLett.117.080601,PhysRevLett.129.094101,PhysRevLett.130.207101}. The study of
EVS has been extremely important in many problems encountered in statistical physics, including disordered systems \cite{fyodorov2008freezing,fyodorov2010multifractality}, fluctuating interfaces \cite{raychaudhuri2001maximal,majumdar2004exact}, interacting
spin systems \cite{bar2016exact}, stochastic transport models \cite{majumdar2010real,guillet2020extreme}, random matrices \cite{dean2006large,majumdar2009large,majumdar2014top}, epidemic outbreak \cite{dumonteil2013spatial}, binary search trees \cite{krapivsky2000traveling} and related computer search algorithms \cite{majumdar2002extreme,majumdar2003traveling}.  We refer the reader to \cite{fortin2015applications,majumdar2020extreme} for two recent reviews on the EVS.

One of the central goals on this subject is to compute
the statistics of extremes, i.e., the maximum $M$ of a given
trajectory $x(t)$ during an observation time window $\left[0, t \right] $, and the time $t_m$ at which the maximum $M$ is reached. A paradigmatic example is the one-dimensional Brownian motion for a fixed duration $t$ starting from the origin. The joint distribution of $M$ and $t_m$ is given by \cite{schehr2010extreme}
\begin{eqnarray}\label{eq0.1}
P_0(M, t_m|t)=\frac{M}{{2\pi D\sqrt {t_m^3\left( {t - {t_m}} \right)} }}{e^{ - {M^2}/4Dt_m}}
\end{eqnarray}
where $D$ the diffusion constant. Integrating $P_0(M,t_m|t)$ over $t_m$ from 0 to $t$, one can get the marginal distribution of $M$,  
\begin{eqnarray}\label{eq0.2}
P_0(M|t)= \frac{1}{\sqrt{\pi D t}} e^{-M^2/4Dt}, \quad M>0, 
\end{eqnarray}
which is the one-sided Gaussian distribution. Integrating $P_0(M,t_m|t)$ over $M$ from 0 to $\infty$, one can obtain the marginal distribution of $t_m$,  
\begin{eqnarray}\label{eq0.3}
P_0(t_m|t)= \frac{1}{\pi \sqrt{t_m(t-t_m)}}, \quad 0\leq t_m \leq t,
\end{eqnarray}
which is often referred to as the ``arcsine law" due to P. L\'evy \cite{Levy1940ArcsineLaw,feller1971introduction,majumdar2007brownian}. The name stems from the fact that the cumulative distribution of $t_m$ reads $F(z) =  \int_0^z {P(t_m )dt_m}  = (2/\pi)\arcsin \sqrt {z/t}$. A counterintuitive aspect of the $U$-shaped distribution Eq.(\ref{eq0.3}) is that its average value $\langle t_m \rangle=t/2$ corresponds to the minimum of the
distribution, i.e., the less probable outcome, whereas values
close to the extrema $t_m=0$ and $t_m=t$ are much more likely.
Recent studies led to many extensions of the law, such as in constrained Brownian motions \cite{majumdar2008time}, random acceleration process \cite{majumdar2010time,boutcheng2016occupation}, fractional Brownian motion \cite{sadhu2018generalized,sadhu2021functionals}, run-and-tumble motion \cite{SinghArcsinelaws_RTP}, resetting Brownian motion \cite{PhysRevE.103.022135,PhysRevE.103.052119}, and for general
stochastic processes \cite{lamperti1958occupation,kasahara1977limit,dhar1999residence,majumdar2002exact,schehr2010extreme,PhysRevLett.107.170601,PhysRevE.83.061146,PhysRevE.105.024113,PhysRevE.102.032103}. Extension to study the distribution of the time difference
between the minimum and the maximum for stochastic processes has also been made in \cite{mori2019time,mori2020distribution}. Quite remarkably, the statistics of $t_m$ has found applications in convex hull problems \cite{randon2009convex,PhysRevE.103.022135} and also in detecting whether a stationary process is equilibrium or not \cite{mori2021distribution,mori2022time}.

While the statistics of $M$ and $t_m$  in a fixed duration time has been extensively studied, the study of these quantities for a stochastic process until a stopping time, e.g. the first passage time when the process arrives at some threshold value brings some recent attention. This problem is relevant to some context. For instance, in queue theory the maximum queue length and the time at which this length is achieved before the queue length gets to zero \cite{randon2007distribution}. In stock market, an agent can hold the stock till its price reaches a certain threshold value. A best time to sell the stock is when the price of the stock reaches its maximum before dropping to the threshold \cite{majumdar2008optimal}. Another example arises in the biological context
regarding the maximal excursions of the tracer proteins before binding at a site \cite{RevModPhys.83.81,PhysRevX.7.011019}. For a one-dimensional Brownian motion starting from a positive position $x_0$, the marginal distributions of $M$ and $t_m$ before its first passage through the origin were studied analytically \cite{randon2007distribution,kearney2005area}. It was shown that $P_0(t_m|x_0)$ exhibits power-law forms at both large and small tails with $P_0(t_m|x_0) \sim t_m^{-1/2}$ as $t_m \to 0$ and $P_0(t_m|x_0) \sim t_m^{-3/2}$ as $t_m \to \infty$. In the presence of a drift towards the origin, the distribution of $M$ was also obtained analytically, but the distribution of $t_m$ has no a compact expression. An asymptotic analysis showed that the distribution of $t_m$ has the same behavior as the drift-less case as $t_m \to 0$, and has an exponential decay as $t_m \to \infty$ \cite{randon2007distribution}. A recent study has extended to compute the joint distribution of $M$ and $t_m$ and their marginal distributions for the run-and-tumble particle in one dimension \cite{singh2022extremal}. For random walks on one-dimensional lattices, the distribution of a related observable of EVS, i.e., the number of distinct sites visited by the walker before hitting a target, and the joint distribution with the first-passage time to the target were obtained analytically \cite{PhysRevE.103.032107,PhysRevE.105.034116}. Recently, the statistics of the random functionals along the first-passage trajectories has attracted a lot of interest and was studied for instance in the case of Brownian motion \cite{kearney2005area,kearney2007first,majumdar2020statistics}, Brownian motion with stochastic resetting \cite{JPA2022.55.234001,DubeyJPA2023}, Brownian motion in the presence of external potential \cite{kearney2021statistics,PhysRevE.108.044151}, and non-Markov stochastic processes \cite{PhysRevE.107.064122,PhysRevE.105.064137,hartmann2023first}.

Stochastic resetting refers to a renewal process in which the
dynamics is interrupted stochastically followed by its starting anew. The subject has recently received  considerable attention due to wide applications in search problems \cite{PhysRevLett.113.220602,PhysRevE.92.052127}, the optimization of randomized computer algorithms \cite{PhysRevLett.88.178701}, and in the field of
biophysics \cite{reuveni2014role,rotbart2015michaelis} (see \cite{evans2020stochastic,Gupta2022Review} for two recent reviews). Evans and Majumdar studied a paradigmatic model in which a one-dimensional Brownian motion is instantaneously reset to its initial position at a constant rate, while it diffuses freely between two consecutive resetting events. The resetting can lead to a wealth of intriguing phenomena. A nonequilibrium stationary state with non-Gaussian fluctuations for the particle position occurs. The mean time to reach a given target for the first time can become finite and be minimized with respect to the resetting rate \cite{evans2011diffusion}. Some extensions have been made in the field, such as spatially \cite{evans2011diffusion2} or temporally \cite{NJP2016.18.033006,pal2016diffusion,PhysRevE.93.060102,PhysRevE.96.012126,PhysRevE.100.032110} dependent resetting rate,  higher dimensions \cite{Evans2014_Reset_Highd}, complex geometries and networks \cite{Christou2015,PhysRevE.99.032123,PhysRevResearch.2.033027,BressloffJSTAT2021,PhysRevE.105.034109,PhysRevE.101.062147,huang2021random,JSM2022.053201,PhysRevE.106.044139}, non-instantaneous resetting \cite{EvansJPA2018,PalNJP2019,PhysRevE.101.052130,gupta2020stochastic,mercado2020intermittent,santra2021brownian,radice2021one}, in the presence of external potential \cite{pal2015diffusion,ahmad2019first,gupta2020stochastic}, other types of Brownian motion, like run-and-tumble particles \cite{evans2018run,santra2020run,bressloff2020occupation}, active particles \cite{scacchi2018mean,kumar2020active}, constrained Brownian particle \cite{PhysRevLett.128.200603}, and so on \cite{basu2019symmetric}.  
The resetting can also find its interesting applications in diverse fields including statistical physics \cite{PhysRevLett.116.170601,pal2017first,gupta2014fluctuating,meylahn2015large,chechkin2018random,PhysRevE.103.022135,de2020optimization,magoni2020ising,arXiv:2202.04906,JPA2022.55.021001,JPA2022.55.234001,sokolov2023linear}, stochastic thermodynamics \cite{fuchs2016stochastic,pal2017integral,gupta2020work}, chemical and biological processes \cite{reuveni2014role,PhysRevLett.112.240601,rotbart2015michaelis,PhysRevLett.128.148301,JPA2022.55.274005}, record statistics \cite{JStatMech2022.063202,JPA2022.55.034002,kumar2023universal,smith2023striking}, optimal control theory \cite{arXiv:2112.11416}, and single-particle experiments \cite{tal2020experimental,besga2020optimal}.

In a recent work \cite{PhysRevE.108.044115}, we have studied the EVS of a resetting Brownian particle in one dimension till it passes through the origin for the first time, starting from a positive position $x_0$. Therein, we have obtained analytically the marginal distribution $P_r(M|x_0)$ of the maximum displacement $M$, and in the Laplace domain the joint distribution $P_r(M,t_m|x_0)$ of $M$ and the time $t_m$ at which $M$ is reached. An interesting observation is that there exists an optimal resetting rate at which the mean extreme time $\langle t_m \rangle$ shows its unique minimum. In the present work, we consider a one-dimensional resetting \textit{drifted} Brownian motion, and study the effect of drift on the EVS.  We find that the introduction of the drift can produce a significant effect on the statistics of $M$ and $t_m$, and induce more abundant behaviors with respect to the resetting rate $r$. The mean value of the maximum displacement $M$ can show three different trends of change with $r$ depending on the drift velocity $v$. For $v \geq 0$ ($v<v_c$), $\langle M \rangle$ decreases (increases) monotonically with $r$. While for $0>v>v_c$ ($v_c \approx -1.69415D/x_0$), $\langle M \rangle$ is a nonmonotonic function of $r$. For the latter, $\langle M \rangle$ has a unique minimum at some nonzero $r$. The mean value $\langle t_m \rangle$ of the extreme time also exhibits nontrivial dependence on $r$. When the drift velocity $v$ is larger than a critical value $v_m \approx -1.66102D/x_0$, there is a nonzero resetting rate at which  $\langle  t_m \rangle$ is a minimum. Otherwise, $\langle  t_m \rangle$ increases monotonically with $r$. This indicates a resetting-induced transition phenomenon for optimizing the mean extreme time.

\section{Model}
Let us consider a one-dimensional resetting drifted Brownian motion (RDBM), starting from $x_0$ and resetting instantaneously to the position $x_r$ at random times but with a constant rate $r$. The position $x(t)$ of the particle at time $t$ is updated by the following stochastic rule
\begin{eqnarray}\label{eq1.0}
{x}( {t + dt} ) = \left\{ \begin{array}{lll}
{x}( t ) + v dt+ \sqrt {2 D } {\xi }(t) dt , &  {\rm{with \ prob.}}    & 1-r dt,  \\
{x}_r, &   {\rm{with \ prob.}}   & r dt,  \\ 
\end{array}  \right. 
\end{eqnarray}
where $v$ is the drift velocity, $D$ is the diffusion constant, and ${\xi }(t)$ is a Gaussian white noise with zero mean $\langle {\xi (t)} \rangle  = 0$ and delta correlator $\langle {\xi (t)}{\xi (t')} \rangle  = \delta ( {t - t'} )$.

\begin{figure}
	\centerline{\includegraphics*[width=0.6\columnwidth]{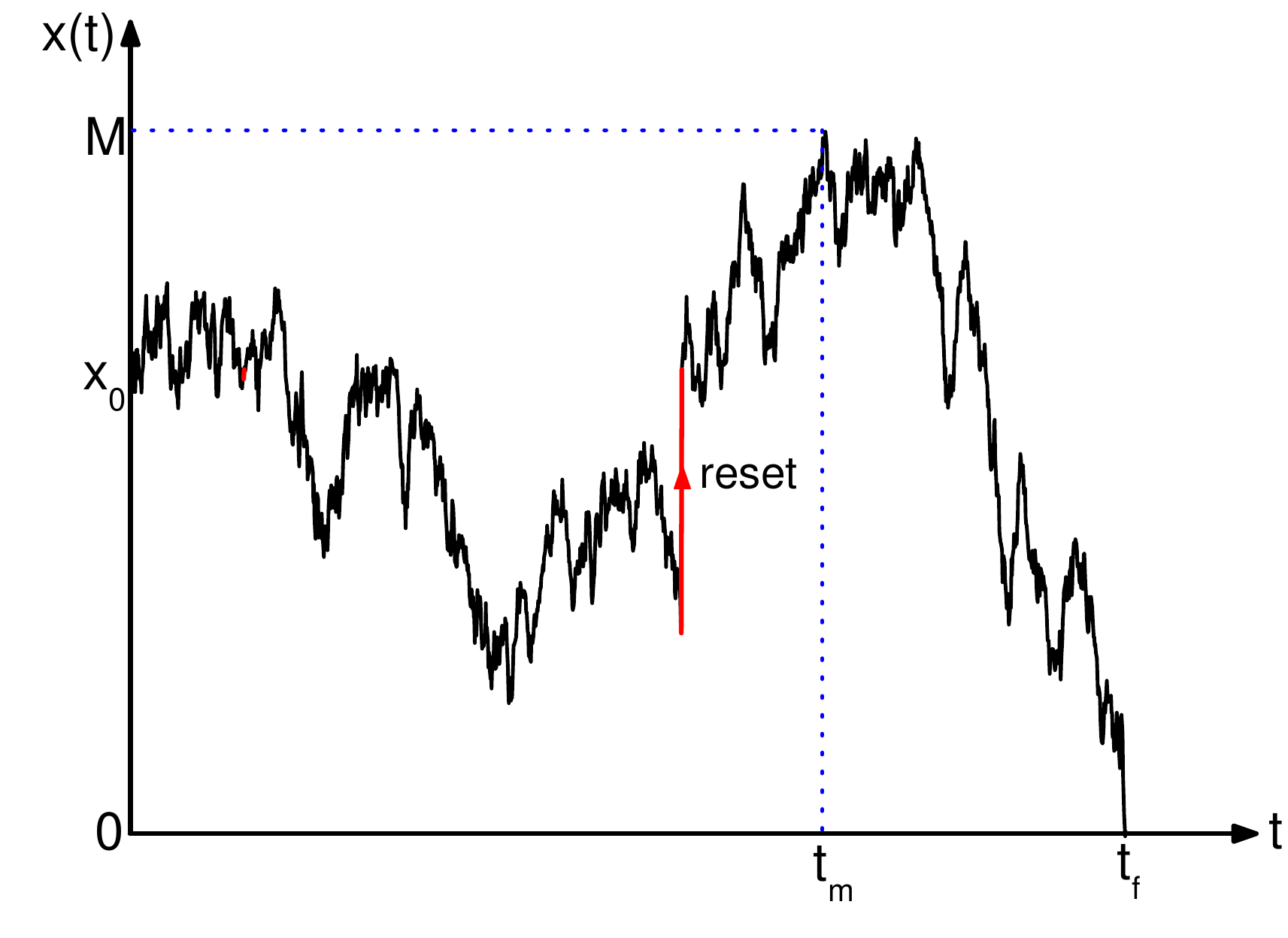}}
	\caption{A realization of a one-dimensional resetting drift Brownian motion in the presence of an absorbing wall at the origin. The process $x(t)$ starts from $x_0$ and reaches its maximum $M$ at time $t_m$ before the first-passage time $t_f$ through the origin. The resetting events are marked by red vertical lines. \label{fig1}}
\end{figure}

We are interested in the extremal statistics of an ensemble of the first-passage trajectories, where each trajectory starts from a position $x_0>0$ and terminates whenever it hits the origin at time $t_f$ for the first time, as shown in Fig.\ref{fig1}. Along the first-passage trajectory, the displacement $x(t)$ reaches its maximum $M$ at time $t_m$. The main purpose of the present work is to compute the joint distribution $P_r(M, t_m|x_0)$ of $M$ and $t_m$, and their marginal distributions, i.e., $P_r(M|x_0)$ and $P_r(t_m|x_0)$.

Besides the extreme time $t_m$, the first-passage time $t_f$ is also fluctuating and varies from one realization to another one. The mean first-passage time $\langle t_f \rangle$ was obtained in a previous work, given by \cite{JPA2019.52.255002}
\begin{eqnarray}\label{eq1.1}
\langle t_f \rangle=\frac{{{e^{\left( {v/2D + {w_{0,r}}} \right){x_0}}} - 1}}{r}
\end{eqnarray}
where \textcolor{blue}{ we have defined the notation for convenience 
\begin{eqnarray}\label{eq1.2}
w_{s,r}=\sqrt{\frac{s+r}{D}+\frac{v^2}{4D^2}}.
\end{eqnarray}
} In the absence of resetting ($r=0$), $\langle t_f \rangle$ is divergent for $v\geq 0$, but converges to $\frac{x_0}{|v|}$ for $v<0$. For $r \to \infty$, $\langle t_f \rangle$ is divergent whether the value of $v$ is. Interestingly, $\langle t_f \rangle$ shows a nonmonotonic dependence on $r$ for $v>v^{*}$, where a unique minimum of $\langle t_f \rangle$ shows up at a nonzero value of $r$. Otherwise, for $v<v^{*}$, $\langle t_f \rangle$ increases monotonically with $r$. Thus, the so-called resetting transition occurs at $v=v^{*}$, where
\begin{eqnarray}\label{eq1.3}
v^{*}=-\frac{2D}{x_0}
\end{eqnarray}
can be obtained by the critical condition: $\frac{\partial \langle t_f \rangle}{\partial r} |_{r=0}=0$. 
  
\section{Survival probability}
Consider a RDBM confining in an interval $\left[0, M \right]$ with the absorbing boundaries at both ends. Let us denote by $Q_r(x_0,t;x_r)$ the survival probability that the particle has touched neither of the boundaries until time $t$ starting from $x_0 \in \left[0, M \right]$. $Q_r(x_0,t;x_r)$ can be connected to its counterpart in the absence of resetting via a renewal formula \cite{chechkin2018random,pal2016diffusion,PhysRevE.99.032123,evans2020stochastic}, 
\begin{equation}\label{eq2.1}
{Q_r}({x_0},t;x_r) = {e^{ - rt}}{Q_0}({x_0},t) + r\int_0^t {d\tau {e^{ -r\tau}}} {Q_0}({x_r},\tau ){Q_r}({x_0},t - \tau ;x_r),
\end{equation}
where $Q_0(x_0,t)$ the survival probability without resetting. Eq.(\ref{eq2.1})
has a simple interpretation. The first term on the right-hand side implies that the particle survives until time $t$ without experiencing any reset event. The second term considers the possibility when there are at least one reset event. One considers that the last reset event has occurred at time $t-\tau$, and after that there has been no reset for the duration $\tau$. This probability is given by $r d\tau e^{-r\tau}$, but then this has to be multiplied by $Q_r(x_0,t-\tau;x_r)$, i.e., the probability that the particle survives until time $t-\tau$ and $Q_0(x_r,t)$, i.e., the survival probability of the particle for the last non-resetting interval $\tau$.

It is convenient to take the Laplace transform of Eq.(\ref{eq2.1}), ${\tilde Q_r}({x_0},s;x_r)=\int_{0}^{\infty} dt e^{-st}{Q_r}({x_0},t;x_r) $, so that the convolution structure in Eq.(\ref{eq2.1}) can be exploited. The Laplace transform of Eq.(\ref{eq2.1}) then satisfies
\begin{equation} \label{eq2.5}
{\tilde Q_r}({x_0},s;x_r) = \frac{{{{\tilde Q}_0}({x_0},s + r)}}{{1 - r{{\tilde Q}_0}({x_r},s + r)}}
\end{equation}
The survival probability ${Q_0}(x_0,t)$ without resetting can be obtained by solving a backward equation, 
\begin{equation}\label{eq2.2}
\frac{{\partial {Q_0}(x_0,t)}}{{\partial t}} = D\frac{{\partial {Q_0}(x_0,t)}}{{\partial {x_0}^2}} + v\frac{{\partial {Q_0}(x_0,t)}}{{\partial {x_0}}},
\end{equation}
subject to the boundary conditions, $Q_0(0,t)=Q_0(M,t)=0$. Performing the Laplace transform for ${Q_0}(x_0,t)$, $\tilde{Q}_0(x_0,s)=\int_{0}^{\infty} dt e^{-st}{Q_0}(x_0,t)$, Eq.(\ref{eq2.2}) becomes
\begin{equation}\label{eq2.3}
s{{\tilde Q}_0}(x_0,s) - 1 = D\frac{{\partial {{\tilde Q}_0}(x_0,s)}}{{\partial {x_0}^2}} + v\frac{{\partial  {{\tilde Q}_0}(x_0,s)}}{{\partial {x_0}}},
\end{equation}
Eq.(\ref{eq2.3}) can be solved in combination with the boundary conditions
\begin{equation}\label{eq2.4}
{{\tilde Q}_0}(0,s) = {{\tilde Q}_0}(M,s) = 0,
\end{equation} 
which yields, 
\begin{equation}\label{eq2.6}
{{\tilde Q}_0}({x_0},s) = \frac{{{{{e}}^{\frac{{ - Mv}}{{2D}}}}\sinh ({w_{s,0}}M)
		- {{{e}}^{\frac{{ - {x_0}v}}{{2D}}}}\sinh ({w_{s,0}}{x_0}) - {{{e}}^{\frac{{ - \left( {M + {x_0}} \right)v}}{{2D}}}}
		\sinh \left[ {{w_{s,0}}\left( {M - {x_0}} \right)} \right]}}{{{{s{e}}^{\frac{{ - Mv}}{{2D}}}}\sinh ({w_{s,0}}M)}},
\end{equation}
where \textcolor{blue}{$w_{s,0}$ is defined in Eq.(\ref{eq1.2})}.

Plugging Eq.(\ref{eq2.6}) into Eq.(\ref{eq2.5}), we obtain the expression of the Laplace transform of $Q_r(x_0,s;x_r)$
\begin{equation}
{{\tilde Q}_r}({x_0},s;x_r) = \frac{{{{{e}}^{\frac{{ - Mv}}{{2D}}}}\sinh ({w_{s,r}}M) - {{{e}}^{\frac{{ - \left( {M + {x_0}} \right)v}}
				{{2D}}}}\sinh \left[ {w_{s,r}\left( {M - {x_0}} \right)} \right] - {{{e}}^{\frac{{ - {x_0}v}}{{2D}}}}\sinh (w_{s,r}{x_0})}}
{{{{{e}}^{\frac{{ - Mv}}{{2D}}}}s\sinh (w_{s,r}M) + {{{e}}^{\frac{{ - \left( {M + {x_r}} \right)v}}{{2D}}}}
		r\sinh \left[ {w_{s,r}\left( {M - {x_r}} \right)} \right] + {{{e}}^{\frac{{ - {x_r}v}}{{2D}}}}r\sinh (w_{s,r}{x_r})}}
\end{equation}
where $w_{s,r}$ is given in Eq.(\ref{eq1.2}). 

\section{Propagator}
Let us denote by $G_r(x,t|x_0;x_r)$ the propagator of a RDBM in an interval $\left[0, M \right] $  with the absorbing boundaries
at both ends, providing that the motion starts from the position $x_0$. One can write a time-dependent equation
for the propagator $G_r(x,t|x_0;x_r)$ using a last renewal formalism \cite{chechkin2018random,pal2016diffusion,PhysRevE.99.032123,evans2020stochastic}
\begin{equation}\label{eq3.1}
{G_r}(x,t|{x_0};x_r) = {e^{ - rt}}{G_0}(x,t|{x_0}) + r\int_0^t {d\tau {e^{ - r\tau}}} {G_0}(x,\tau |{x_r}){Q_r}({x_0},t - \tau;x_r ) ,
\end{equation}
where $G_0(x,t|x_0)$ is the propagator in the absence of resetting.  Also, we recall that the survival probability until time $t$ is defined as
\begin{equation}\label{eq3.2}
{Q_r}({x_0},t;x_r) = \int_0^M dx {G_r}(x,t|x_0;x_r) .
\end{equation}
By taking the Laplace transform on both sides of Eq.(\ref{eq3.1}) and using Eq.(\ref{eq2.5}), we find
\begin{equation}\label{eq3.3}
{\tilde G_r}(x,s|{x_0};x_r) = {\tilde G_0}(x,s + r|{x_0}) + r{\tilde G_0}(x,s + r|{x_r})
\frac{{{{\tilde Q}_0}({x_0},s + r)}}{{1 - r{{\tilde Q}_0}({x_r},s + r)}} .
\end{equation}
The propagator $\tilde G_0(x,s|x_0)$ is a classical result [see, e.g., Eq.(2.2.28) in Ref.\cite{redner2001guide}] 
\begin{equation}\label{eq3.4}
{\tilde G_0}(x,s|{x_0}) = {e^{\frac{{v(x - {x_0})}}{{2D}}}}\frac{{\cosh [(M - |x - {x_0}|){w_{s,0}}] - \cosh [(M - x - {x_0}){w_{s,0}}]}}{{2D w_{s,0} \sinh (w_{s,0} M)}}.
\end{equation}
Substituting Eq.(\ref{eq2.6}) and Eq.(\ref{eq3.4}) into Eq.(\ref{eq3.3}), we obtain
\begin{eqnarray}\label{eq3.5}
{\tilde G_r}(x,s|{x_0};x_r) &=&  {e^{\frac{{v(x - {x_0})}}{{2D}}}}\frac{{\cosh [(M - |x - {x_0}|) {w_{s,r}} ] - \cosh [(M - x - {x_0})w ]}}{{2D\omega \sinh (w_{s,r} M)}} \nonumber \\ &
+& r{e^{\frac{{v(x - {x_r})}}{{2D}}}}\frac{{\cosh [(M - |x - {x_r}|)w_{s,r} ] - \cosh [(M - x - {x_r})w_{s,r} ]}}{{2D w_{s,r} \sinh (w_{s,r}M)}} \nonumber \\ &\times&
\frac{{{{{e}}^{\frac{{ - Mv}}{{2D}}}}\sinh (w_{s,r}M) - {{{e}}^{\frac{{ - \left( {M + {x_0}} \right)v}}	{{2D}}}}\sinh \left[ {w_{s,r}\left( {M - {x_0}} \right)} \right] - {{{e}}^{\frac{{ - {x_0}v}}{{2D}}}}\sinh (w_{s,r}{x_0})}}
{{{{{e}}^{\frac{{ - Mv}}{{2D}}}}s\sinh (w_{s,r}M) + {{{e}}^{\frac{{ - \left( {M + {x_r}} \right)v}}{{2D}}}}	r\sinh \left[ {w_{s,r}\left( {M - {x_r}} \right)} \right] + {{{e}}^{\frac{{ - {x_r}v}}{{2D}}}}r\sinh (w_{s,r}{x_r})}}.
\end{eqnarray}
For the special case, $x_r$=$x_0$, Eq.(\ref{eq3.5}) simplifies to
\begin{eqnarray}\label{eq3.6}
{{\tilde G}_r}(x,s|{x_0}) =\frac{{\left( {r + s} \right)}}{{2Dw_{s,r}}}\frac{{{{{e}}^{\frac{{v\left( {M + x + {x_0}} \right)}}
				{{2D}}}}\left( {\cosh \left[ {w_{s,r}\left( {M - |x - {x_0}|} \right)} \right] - \cosh \left[ {w_{s,r}\left( {M - x - {x_0}} \right)} 
			\right]} \right)}}{{{{{e}}^{\frac{{v\left( {M + 2{x_0}} \right)}}{{2D}}}}s\sinh (Mw_{s,r}) + {{{e}}^{\frac{{v\left( {M + {x_0}} \right)}}
				{{2D}}}}r\sinh \left[ {w_{s,r}\left( {M - {x_0}} \right)} \right] + {{{e}}^{\frac{{v\left( {2M + {x_0}} \right)}}{{2D}}}}r\sinh (w_{s,r}{x_0})}},
\end{eqnarray}
where we have dropped the notation $x_r$ in ${{\tilde G}_r}(x,s|{x_0})$ as long as $x_r=x_0$.

\section{Exit probability}
Consider a RDBM starting from $x_0 \in \left[ 0, M \right] $ and both of the interval are absorbing boundaries. Let us denote
by $\mathcal{E}(x_0;x_r)$ the splitting or exit probability that the particle
exits the interval for the first time through the origin, i.e., the probability that the maximum before the first-passage time is less than or equal to $M$. To compute $\mathcal{E}(x_0;x_r)$, it is best to first measure the probability current through the origin,
\begin{equation}\label{eq4.1}
J(t;x_0,x_r) =   D\frac{{\partial {G_r}(x,t|{x_0};x_r)}}{{\partial x}}{|_{x = 0}} ,
\end{equation}
and ${\tilde{J}}(s;x_0,x_r) = \int_0^\infty  {dt} e^{-st} J(t;x_0,x_r)$ satisfies
\begin{equation}\label{eq4.2}
{\tilde{J}}(s;x_0,x_r) = D\frac{{\partial {\tilde{G}_r}(x,s|{x_0};x_r)}}{{\partial x}}{|_{x = 0}}  .
\end{equation}
Using Eq.(\ref{eq3.5}), it follows as
\begin{eqnarray}\label{eq4.3}
{\tilde{J}}(s;x_0,x_r)&=&  {e^{\frac{{ - {x_0}v}}{{2D}}}}\frac{{\sinh \left[ {w_{s,r}\left( {M - {x_0}} \right)} \right]}}{{\sinh (w_{s,r}M)  }} + r{e^{\frac{{ - {x_r}v}}{{2D}}}} \nonumber \\ &\times& \frac{{\sinh \left[ {w_{s,r}\left( {M - {x_r}} \right)} \right]}}{{\sinh (w_{s,r}M)}}
\frac{{{{{e}}^{\frac{{ - Mv}}{{2D}}}}\sinh (w_{s,r}M) - {{{e}}^{\frac{{ - \left( {M + {x_0}} \right)v}}
				{{2D}}}}\sinh \left[ {w_{s,r}\left( {M - {x_0}} \right)} \right] - {{{e}}^{\frac{{ - {x_0}v}}{{2D}}}}\sinh (w_{s,r}{x_0})}}
{{{{{e}}^{\frac{{ - Mv}}{{2D}}}}s\sinh (w_{s,r}M) + {{{e}}^{\frac{{ - \left( {M + {x_r}} \right)v}}{{2D}}}}
		r\sinh \left[ {w_{s,r}\left( {M - {x_r}} \right)} \right] + {{{e}}^{\frac{{ - {x_r}v}}{{2D}}}}r\sinh (w_{s,r}{x_r})}},
\end{eqnarray}
$\mathcal{E}(x_0;x_r)$ can be obtained by integrating the probability current through the origin over the time, 
\begin{eqnarray}\label{eq4.4}
\mathcal{E}(x_0;x_r) = \int_0^\infty  {dt} J(t;x_0,x_r) = {\tilde{J}}(s=0;x_0,x_r).
\end{eqnarray}
Thus, we get the exit probability from the origin, 
\begin{eqnarray}\label{eq4.5}
\mathcal{E}(x_0;x_r)= \frac{{\sinh \left[ {{w_{0,r}}\left( {M - {x_r}} \right)} \right] 
		+ {{{e}}^{\frac{{v\left( {M - {x_0}} \right)}}{{2D}}}}\sinh \left[ {{w_{0,r}}\left( {{x_r} - {x_0}} \right)} \right]}}{{\sinh \left[ {{w_{0,r}}\left( {M - {x_r}} \right)} \right] + {{{e}}^{\frac{{Mv}}{{2D}}}}\sinh ({w_{0,r}}{x_r})}},
\end{eqnarray}
where $w_{0,r}$ is given in Eq.(\ref{eq1.2}). For $x_r=x_0$, Eq.(\ref{eq4.5}) reduces to
\begin{equation}\label{eq4.7}
\mathcal{E}(x_0) = \frac{{\sinh \left[ {{w_{0,r}}\left( {M - {x_0}} \right)} \right]}}{{\sinh \left[ {{w_{0,r}}\left( {M - {x_0}} \right)} \right] + {{{e}}^{\frac{{Mv}}{{2D}}}}\sinh ({w_{0,r}}{x_0})}}.
\end{equation}
where we have suppressed the notation $x_r$ in $\mathcal{E}(x_0)$ whenever $x_r=x_0$. \textcolor{blue}{For a fixed $M$, $\mathcal{E}(x_0)$ is a monotonically decreasing function of $x_0$, which varies from $\mathcal{E}(x_0)=1$ at $x_0=0$ to $\mathcal{E}(x_0)=0$ at $x_0=M$. In the limits of $x_0 \to 0^{+}$ and $x_0 \to M^{-}$, one has
\begin{equation}\label{eq4.7.0}
\mathcal{E}(x_0) \sim \left\{ \begin{array}{ll}
1 - \frac{{{w_{0,r}}{e^{\frac{{Mv}}{{2D}}}}}}{{\sinh \left( {{w_{0,r}}M} \right)}}{x_0},  &x_0\to 0^{+}, \\
\frac{{{w_{0,r}}{e^{ - \frac{{Mv}}{{2D}}}}}}{{\sinh \left( {{w_{0,r}}M} \right)}}\left( {M - {x_0}} \right),  &x_0\to M^{-} .
\end{array}  \right. 
\end{equation}}

\section{Distribution of the maximum displacement $M$}
As mentioned before, the exit probability from the origin corresponds to the probability that the maximum displacement before the first-passage time is less than or equal to $M$. Therefore, differentiating Eq.(\ref{eq4.7}) with respect to $M$ gives the probability density of $M$, 
\begin{equation}\label{eq5.1}
P_r(M|x_0) = \frac{\partial \mathcal{E}(x_0)}{\partial M}=\frac{{{e^{\frac{{vM}}{{2D}}}}\sinh \left( {{w_{0,r}}{x_0}} \right)\left( {2D{w_{0,r}}\cosh \left[ {{w_{0,r}}\left( {M - {x_0}} \right)} \right] - v\sinh \left[ {{w_{0,r}}\left( {M - {x_0}} \right)} \right]} \right)}}{{2D{{\left( {\sinh \left[ {{w_{0,r}}\left( {M - {x_0}} \right)} \right] + {e^{\frac{{vM}}{{2D}}}}\sinh \left( {{w_{0,r}}{x_0}} \right)} \right)}^2}}}.
\end{equation}

\begin{figure}
	\centerline{\includegraphics*[width=0.6\columnwidth]{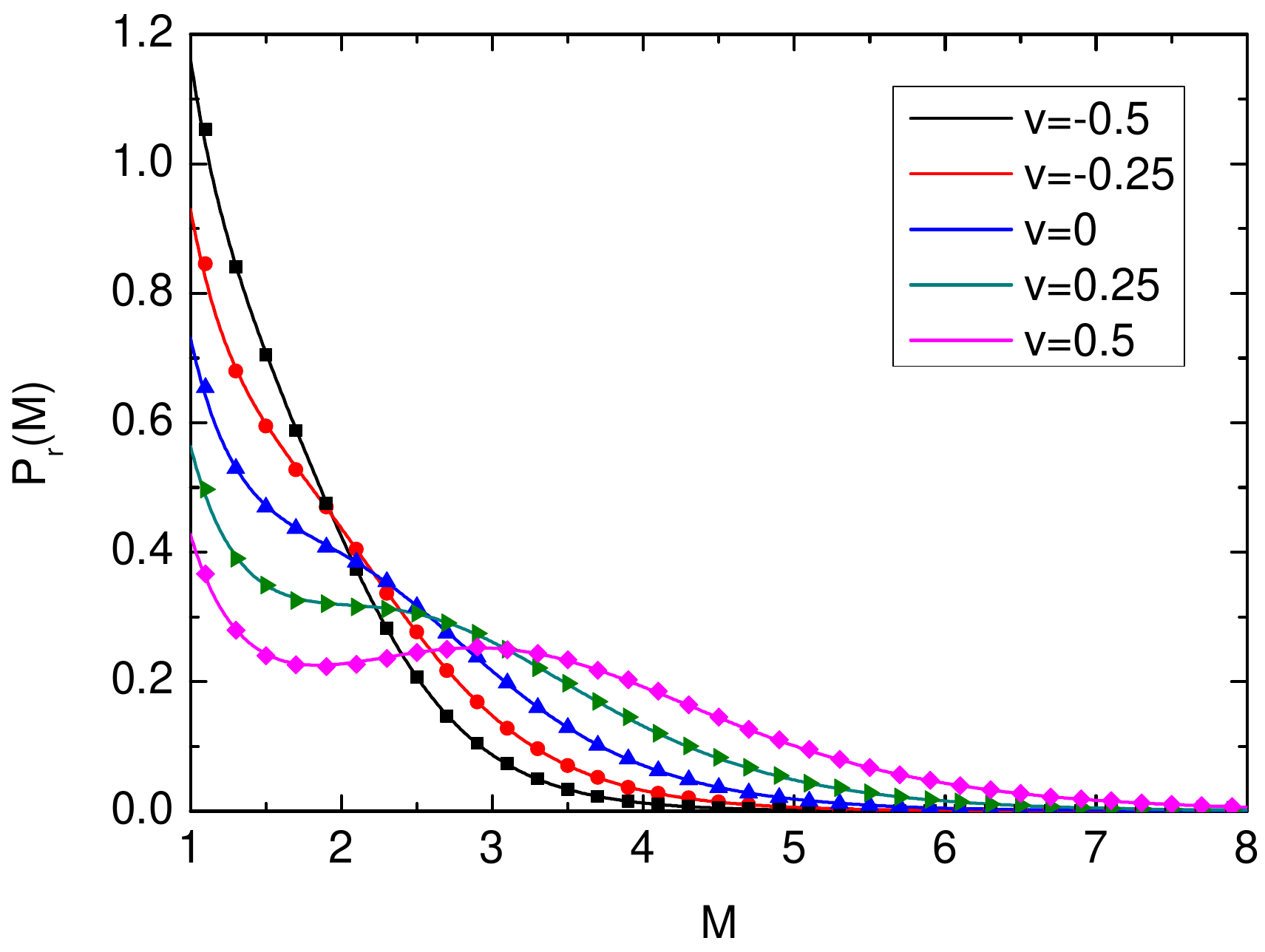}}
	\caption{The marginal distribution $P_r(M|x_0)$ of the maximum displacement $M$ for different drift velocity $v$, where $r=1$, $x_0=1$ and $D=1/2$ are fixed. The lines and symbols correspond to the theoretical and simulation results, respectively. \label{fig2}}
\end{figure}

\textcolor{blue}{ In the limit of $r \to 0$, Eq.(\ref{eq5.1}) reduces to the result without resetting 
\begin{equation}\label{eq5.1.0}
P_0(M|x_0) =\frac{{v\sinh \left( {\frac{{{x_0}v}}{{2D}}} \right)}}{{2D{{\sinh }^2}\left( {\frac{{Mv}}{{2D}}} \right)}}{e^{ - \frac{{{x_0}v}}{{2D}}}},
\end{equation}
which is consistent with the known result \cite{randon2007distribution,kearney2005area}.  
For any nonzero resetting rate $r$, $P_r(M|x_0)$ decays exponentially with $M$ in the large-$M$ limit, $P_r(M|x_0) \sim e^{-({w_{0,r}}-v/2D) M}$. }

In Fig.\ref{fig2}, we show $P_r(M|x_0)$ for different values of $v$, where we have fixed $r=1$, $x_0=1$ and $D=1/2$. We also perform the extensive numerical simulations, as shown by symbols in Fig.\ref{fig2}. In the simulations, we have used a time step $dt=10^{-4}$ and each data is obtained by averaging over $10^5$ first-passage trajectories. Clearly, the simulations agree well with the theory. \textcolor{blue}{Interestingly, as $v$ increases $P_r(M|x_0)$ changes from a simple decay as $M$ to a nonmonotonic function of $M$. There exists a critical drift at which the transition happens. The critical point can be obtained by the conditions: $\partial P_r(M|x_0) / \partial M=0$ and $\partial^2 P_r(M|x_0) / \partial M^2=0$, just as determining the critical point of van der Waals equation. However, we are unable to analytically obtain the critical point. Instead, numerical solution shows that the critical point $\left(v, M \right)  \approx \left(0.304, 2.1 \right)$, where the other parameters are the same as those in Fig.\ref{fig2}.          }

The expected value of the maximum displacement $M$ is computed as
\begin{equation}\label{eq5.2}
\langle M(x_0) \rangle  = \int_{{x_0}}^\infty  {dM} M{P_r}(M|x_0) .
\end{equation}
However, \textcolor{blue}{we are unable to perform the integration in Eq.(\ref{eq5.2}) analytically}, except for the specific case $v=0$ where the expression of $\langle M(x_0) \rangle$ was obtained in one of our recent works \cite{PhysRevE.108.044115}.

Let us first consider two limiting cases. \textcolor{blue}{In the absence of resetting, $r=0$, the expectation of $M$ is divergent for $v \geq 0$. This is because that the particle will escape to infinity with a nonzero probability without touching the absorbing boundary at the origin \cite{redner2001guide}. This means that with a finite probability $t_f \to \infty$ and thus the expectation of $M$ is divergent. While for $v<0$, the particle will eventually reach zero with probability one, and therefore the expectation of $M$ is  convergent. }
The expectation of $M$ without resetting is given by
\begin{equation}\label{eq5.4}
\langle M(x_0) \rangle_{r=0}=\left\{ \begin{array}{lll}
+\infty, &{\rm{for}} &v\geq 0, \\
{x_0} + \frac{D}{v}\left( {{e^{ - \frac{{v{x_0}}}{D}}} - 1} \right)\ln \left( {1 - {e^{\frac{{v{x_0}}}{D}}}} \right),  &{\rm{for}} &v<0 .
\end{array}  \right. 
\end{equation}
\textcolor{blue}{As $v \to 0^{-}$, the expectation of $M$ is divergent logarithmically, $\langle M(x_0) \rangle_{r=0} \sim -x_0 \ln\left(\frac{|v| x_0}{D} \right)$. This is in contrast with the power-law diveregnce of the mean first-passage time $\langle t_f \rangle \sim |v|^{-1}$ \cite{redner2001guide}. }

In the limit of $r\to \infty$, the exit probability $\mathcal{E}(x_0)$ becomes a step-like function, i.e., $\mathcal{E}(x_0)=1$ for $x_0<M/2$ and $\mathcal{E}(x_0)=0$ for $x_0>M/2$. Taking the derivative of $\mathcal{E}(x_0)$ with respect to $M$, we obtain the distribution of $M$, which is a delta function, $P_{r \to \infty}(M|x_0)=\delta(M-2x_0)$. Therefore, the expectation of $M$ in the limit of $r\to \infty$ is given by
\begin{equation}\label{eq5.5}
\langle M(x_0) \rangle=2x_0, \quad {\rm{as} } \quad r \to \infty .
\end{equation}

\begin{figure}
	\centerline{\includegraphics*[width=1.0\columnwidth]{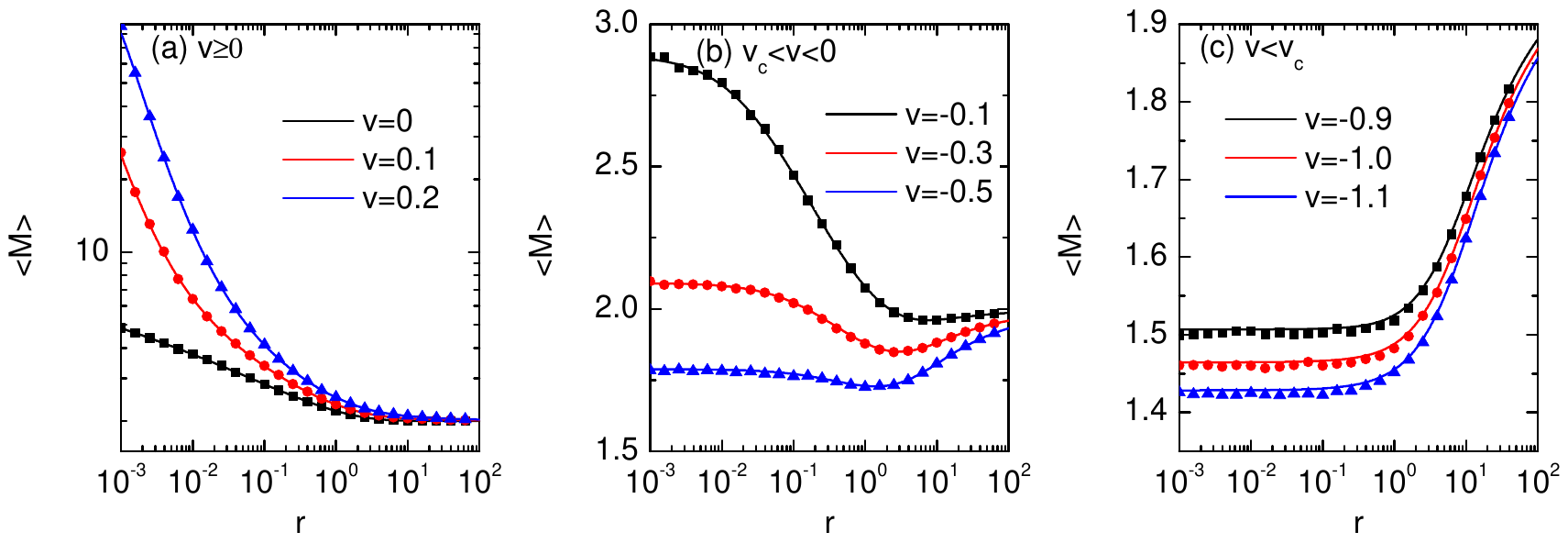}}
	\caption{The expected value $\langle M(x_0) \rangle$ of the maximum displacement $M$ as a function the resetting rate $r$ for several different values of $v$, where $x_r=x_0=1$ and $D=1/2$. The lines  and symbols correspond to the theoretical and simulation results, respectively. \label{fig3}}
\end{figure}

In Fig.\ref{fig3}, we plot $\langle M(x_0) \rangle$ as a function of $r$ for different values of $v$. One can observe that the dependence of $\langle M(x_0) \rangle$  on $r$ shows three distinct regions depending on the value of $v$. 
\begin{itemize}
	\item For $v \geq 0$, $\langle M(x_0) \rangle$ decreases monotonically with $r$.
	\item For $0>v>v_c$, $\langle M(x_0) \rangle$ is a nonmonotonic function of $r$, where $v_c$ is negative and will be determined later. Under this case, there is some nonzero $r$ at which $\langle M(x_0) \rangle$ shows its unique minimum. 
	\item For $v<v_c$, $\langle M(x_0) \rangle$ increases monotonically with $r$.
\end{itemize}

\textcolor{blue}{The EVS of the first-passage trajectories can be related to many practical situations. For example, during the process of animal foraging the range of animal activity before successful foraging is measured by the maximum displacement. Our results show that the resetting plays a nontrivial role in the process when a bias in the direction of movement is present. A particular intriguing case occurs when the drift velocity $v$ points to the target and the amplitude of the drift is less than a critical value $|v_c|$. There exists an intermediate level of resetting rate at which the mean maximum displacement can be minimized. Thus, our result may provide a potential application for designing an optimal foraging strategy to minimize the range of animal activity in successful foraging by stochastic resetting. }

We now turn to the determination of the value of $v_c$. It is not hard to see that at $v=v_c$ the deriative of $\langle M(x_0) \rangle$ with respect to $r$ at $r=0$ changes its sign. Under the critical condition, we find that $v_c$ satisfies the following equation, 
\begin{equation}\label{eq5.6}
- ( {1 + y_c} )\left[ {y_c\ln y_c + ( {1 - y_c + \ln y_c} )\ln ( {1 - y_c} )} \right] + ( {y_c - 1} ){\rm{Li}}_2(y_c)=0 ,
\end{equation}
where $y_c={{e^{\frac{{v_c{x_0}}}{D}}}} \in \left(0, 1 \right) $ and ${\rm{Li}}_2(\cdot)$ is the polylogarithm function. Eq.(\ref{eq5.6}) can be solved numerically to obtain $y_c = 0.183756... $, or equivalently, 
\begin{equation}\label{eq5.7}
v_c=\ln y_c \frac{D}{x_0} \approx -1.69415 \frac{D}{x_0}
\end{equation}

\section{Joint distribution of the maximum displacement $M$ and the time $t_m$ at which $M$ is reached}
For a typical first-passage trajectory starting from $x_0$ ($>0$) and passing through the origin at time $t_f$ for the first time, the maximum displacement $M$ is reached at time $t_m$. Let us denote by $P_r(M,t_m|x_0)$ the joint distribution of $M$ and $t_m$. To compute $P_r(M,t_m|x_0)$,  one can decompose the trajectory into two
parts: a left-hand segment for which $0<t<t_m$, and a right-hand segment for which $t_m<t <t_f$, as shown in Fig.\ref{fig1}. Due to the Markovian property, the weights of the left and the right segments become completely independent and the total weight is just proportional to the product of the weights of the two separate segments. For the first segment, we have a process that propagates from $x_0$ at $t=0$ to $M$ at $t=t_m$ without touching the origin. The statistical weight of the first segment thus equals to the propagator $G_r(M,t_m|x_0)$.
However, it turns out that $G_r(M,t_m|x_0)=0$ which implies that the contribution
from this part is zero. To circumvent this problem, we compute $G_r(M-\epsilon,t_m|x_0)$ and later take the limit $\epsilon \to 0^{+}$ \cite{majumdar2004exact}. For the second segment, the process propagates
from $M-\epsilon$ at $t_m$ to 0 at $t_f$, where $t_f\geq t_m$ without crossing the level $M$ and the level 0 in between. The statistical weight of the second segment is given by the exit probability $\mathcal{E}(M-\epsilon;x_0)$ \cite{randon2007distribution}. Therefore, the joint probability density $P_r(M,t_m|x_0)$ can be written as the product of the statistical weights of two segments 
\begin{eqnarray}\label{eq6.1}
{P_r}( {M,{t_m}|{x_0}} ) = \lim_{\epsilon \to 0^{+}} \mathcal{N}{G_r}( {M - \epsilon,{t_m}|{x_0}} ){\mathcal{E}}( {M - \epsilon}; x_0 ), 
\end{eqnarray}
where the normalization factor $\mathcal{N}$  will be determined later. It is useful to perform the Laplace transform for $P_r(M,t_m|x_0)$ with respect to $t_m$, 
\begin{eqnarray}\label{eq6.2}
\tilde{P}_r(M,\mu|x_0)=\int_{0}^{\infty} d t_m e^{-\mu t_m} P_r(M,t_m|x_0) = \mathcal{N}{\tilde{G}_r}( {M - \epsilon,{\mu}|{x_0}} ){\mathcal{E}}( {M - \epsilon}; x_0), 
\end{eqnarray} 
where ${\tilde{G}_r}( {M - \epsilon,{\mu}|{x_0}} )$ and ${\mathcal{E}}( {M - \epsilon}; x_0)$ can be obtained from Eq.(\ref{eq3.6}) and Eq.(\ref{eq4.5}). In the leading term in $\epsilon$, they are 
\begin{eqnarray}\label{eq6.3}
{\tilde{G}_r}( {M - \epsilon,{\mu}|{x_0}} )=\frac{{r + \mu}}{D}\frac{{{e^{\frac{{vM}}{{2D}}}}\sinh \left( {w_{\mu,r}{x_0}} \right)}}{{\mu{e^{\frac{{v{x_0}}}{{2D}}}}\sinh \left( {w_{\mu,r}M} \right) + r\sinh \left[ {w_{\mu,r}\left( {M - {x_0}} \right)} \right] + r{e^{\frac{{vM}}{{2D}}}}\sinh \left( {w_{\mu,r}{x_0}} \right)}} \epsilon,
\end{eqnarray}
and
\begin{eqnarray}\label{eq6.4}
{\mathcal{E}}( {M - \epsilon}; x_0)=\frac{{2D{w_0}\cosh \left[ {{w_{0,r}}\left( {M - {x_0}} \right)} \right] - v\sinh \left[ {{w_{0,r}}\left( {M - {x_0}} \right)} \right]}}{{2D\sinh \left[ {{w_{0,r}}\left( {M - {x_0}} \right)} \right] + 2D{e^{\frac{{vM}}{{2D}}}}\sinh \left( {{w_{0,r}}{x_0}} \right)}} \epsilon.
\end{eqnarray}

Letting $\mu \to 0$, the left-hand side of Eq.(\ref{eq6.2}) is just the marginal distribution $P_r(M|x_0)$, which yields
\begin{eqnarray}\label{eq6.5}
{P_r}\left( {M|{x_0}} \right) = \int_0^\infty d{t_m}{P_r}\left( {M,{t_m}|{x_0}} \right) =  \mathcal{N}{{\tilde G}_r}\left( {M - \epsilon,0|{x_0}} \right){\mathcal{E}_r}\left( {M - \epsilon} ;x_0\right).
\end{eqnarray}
Plugging Eq.(\ref{eq6.3}) and Eq.(\ref{eq6.4})  into Eq.(\ref{eq6.5}), and then comparing with Eq.(\ref{eq5.1}), we obtain 
\begin{eqnarray}\label{eq6.6}
\mathcal{N}=\frac{D}{\epsilon^2}.
\end{eqnarray}

\begin{figure}
	\centerline{\includegraphics*[width=0.8\columnwidth]{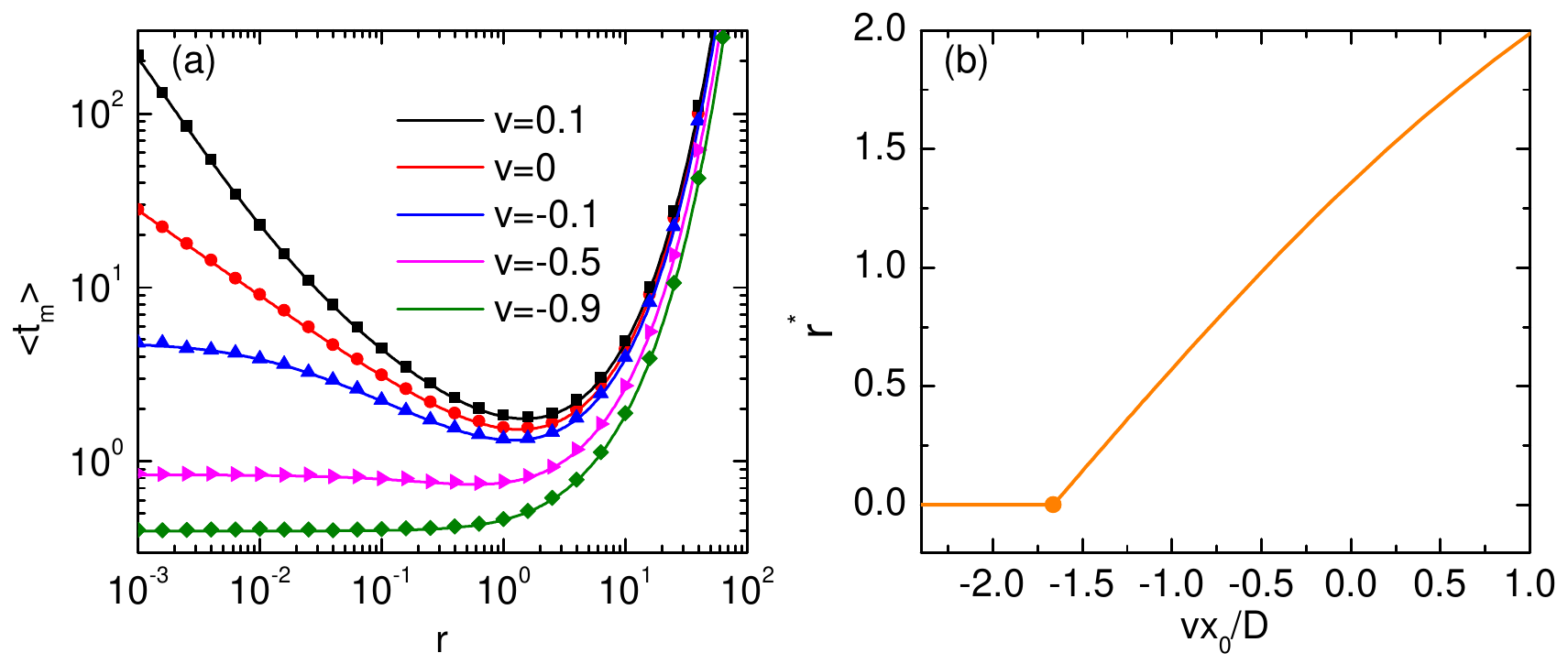}}
	\caption{(a) The expected value $\langle t_m \rangle$ of the time $t_m$ at which the RDBM reaches its maximum before its first passage through the origin, having starting from $x_0>0$, as a function of the resetting rate $r$, where $x_0=1$ and $D=1/2$ are fixed. (b) The optimal resetting rate $r^{*}$ as a function of drift velocity $v$. The lines and symbols correspond to the theoretical and simulation results, respectively. \label{fig4}}
\end{figure}

Finally, we insert Eq.(\ref{eq6.3}), Eq.(\ref{eq6.4}) and Eq.(\ref{eq6.6}) into Eq.(\ref{eq6.2}) to obtain the joint distribution in the Laplace domain, 
\begin{eqnarray}\label{eq6.7}
\tilde{P}_r(M,\mu|x_0)=&& \frac{{\left( {r + \mu} \right){e^{\frac{{vM}}{{2D}}}}\sinh \left( {{w_{\mu,r}}{x_0}} \right)}}{{\mu{e^{\frac{{v{x_0}}}{{2D}}}}\sinh \left( {{w_{\mu,r}}M} \right) + r\sinh \left[ {{w_{\mu,r}}\left( {M - {x_0}} \right)} \right] + r{e^{\frac{{vM}}{{2D}}}}\sinh \left( {{w_{\mu,r}}{x_0}} \right)}}  \nonumber \\ & \times &\frac{{2D{w_{0,r}}\cosh \left[ {{w_{0,r}}\left( {M - {x_0}} \right)} \right] - v\sinh \left[ {{w_{0,r}}\left( {M - {x_0}} \right)} \right]}}{{2D\sinh \left[ {{w_{0,r}}\left( {M - {x_0}} \right)} \right] + 2D{e^{\frac{{vM}}{{2D}}}}\sinh \left( {{w_{0,r}}{x_0}} \right)}}.
\end{eqnarray} 
To obtain the joint distribution $P_r(M,t_m|x_0)$, one has to perform the inverse Laplace transformation for Eq.(\ref{eq6.7}) with respect to $\mu$. Unfortunately, it turns out to be a challenging task. However, the expectation  of the time $t_m$ is given by
\begin{eqnarray}\label{eq6.8}
\langle t_m \rangle=-\lim_{\mu \to 0} \frac{\partial \tilde{P}_r(\mu|x_0)}{\partial \mu},
\end{eqnarray}
where
\begin{eqnarray}\label{eq6.9}
\tilde{P}_r(\mu|x_0)=\int_{x_0}^{\infty} dM  \tilde{P}_r(M,\mu|x_0)
\end{eqnarray}
is the marginal distribution of $t_m$ in the Laplace domain. The integration in Eq.(\ref{eq6.9}) cannot be done analytically except for the case $v=0$ \cite{PhysRevE.108.044115}. In Fig.\ref{fig4}(a), we plot $\langle t_m \rangle$ as a function of $r$ for different values of $v$, where we have fixed other parameters: $x_0=1$ and $D=1/2$. The lines and symbols correspond to the theoretical and simulation results, respectively. There are in good agreements between them. \textcolor{blue}{In the absence of resetting ($r=0$), the diffusing particle is transient for $v \geq0 $, implying that there is a finite probability that the particle can diffuse indefinitely in the positive direction without touching the origin. Therefore, $\langle t_m \rangle$ is divergent for $v \geq0 $. While for $v<0$, the diffusing particle is recurrent, and thus $\langle t_m \rangle$ is convergent, given by} 
\begin{equation}\label{eq6.10}
\langle t_m \rangle_{r=0}=\left\{ \begin{array}{lll}
+\infty, &{\rm{for}} &v\geq 0, \\
\frac{D}{{{v^2}}} + \frac{{{x_0}}}{v}\frac{{{e^{v{x_0}/D}}}}{{1 - {e^{v{x_0}/D}}}},  &{\rm{for}} &v<0 .
\end{array}  \right. 
\end{equation}
\textcolor{blue}{As $v \to 0^{-}$, the expectation of $t_m$ diverges in a power-law way, $\langle t_m \rangle_{r=0} \sim |v|^{-1}$, which shares the same divergence as the mean first-passage time \cite{redner2001guide}. }

In the opposite limit, $r \to \infty$, $\langle t_m \rangle$ is divergent for arbitrary value of $v$. This is due to the divergence of the mean first-passage time for $r \to \infty$. Interestingly, depending on the value of $v$, $\langle t_m \rangle$ exhibits significantly different trends of change with $r$. When $v$ is larger than a critical value,  $v=v_m$, $\langle t_m \rangle$ varies nonmonotonically with $r$, where a minimal $\langle t_m \rangle$ occurs at an optimal value of resetting rate, $r=r^{*}$. Otherwise,  $\langle {t_m} \rangle$ increases \textcolor{blue}{monotonically} with $r$, implying that the optimal resetting rate is always zero for $v<v_m$.

The critical drift velocity $v_m$ can be determined by the condition $\frac{\partial \langle t_m \rangle}{\partial r}|_{r=0}=0$. After cumbersome calculations, we find that $v_m$ satisfies the following equation, 
\begin{eqnarray}\label{eq6.11}
6 y_m + {y_m^2}\left( {2\ln y_m - {{\ln }^2}y_m - 2} \right) + {y_m^3}\left( {4\ln y_m - 6{{\ln }^2}y_m - 14} \right) + {y_m^4}\left( {10 - 6\ln y_m - {{\ln }^2}y_m} \right) && \nonumber \\ + 4{( {1 - y_m} )^4}\ln ( {1 - y_m} )   - 2{( {1 - y_m} )^3}( {1 + y_m} ){\rm{Li}}_2( y_m )=0,&&
\end{eqnarray}
where $y_m=e^{v_m x_0/D} \in\left(0, 1 \right) $.  Numerically solving Eq.(\ref{eq6.11}) to yield $y_m = 0.189945... $, or equivalently, 
\begin{equation}\label{eq6.12}
v_m=\ln y_m \frac{D}{x_0} \approx -1.66102 \frac{D}{x_0}.
\end{equation}

In Fig.\ref{fig4}(b), we show the optimal resetting rate $r^{*}$ as a function of $vx_0/D$. Obviously, a transition occurs at $v=v_m$ above which the optimal resetting rate $r^{*}$ changes from zero to a nonzero value. This is reminiscent of the resetting transition phenomenon in the mean first-passage time \cite{PhysRevLett.116.170601,pal2017first,PhysRevResearch.1.032001,JPA2019.52.255002,PhysRevE.99.032123,JPA2022.55.021001,PhysRevE.105.034109}. Therefore, our finding provide an additional resetting-induced transition phenomenon for optimizing the mean extreme time in the drift-diffusion model. However, we should emphasize that the critical drift velocities for the two mean times are different, see Eq.(\ref{eq1.3}) and Eq.(\ref{eq6.12}) for a comparison.

\begin{figure}
	\centerline{\includegraphics*[width=0.6\columnwidth]{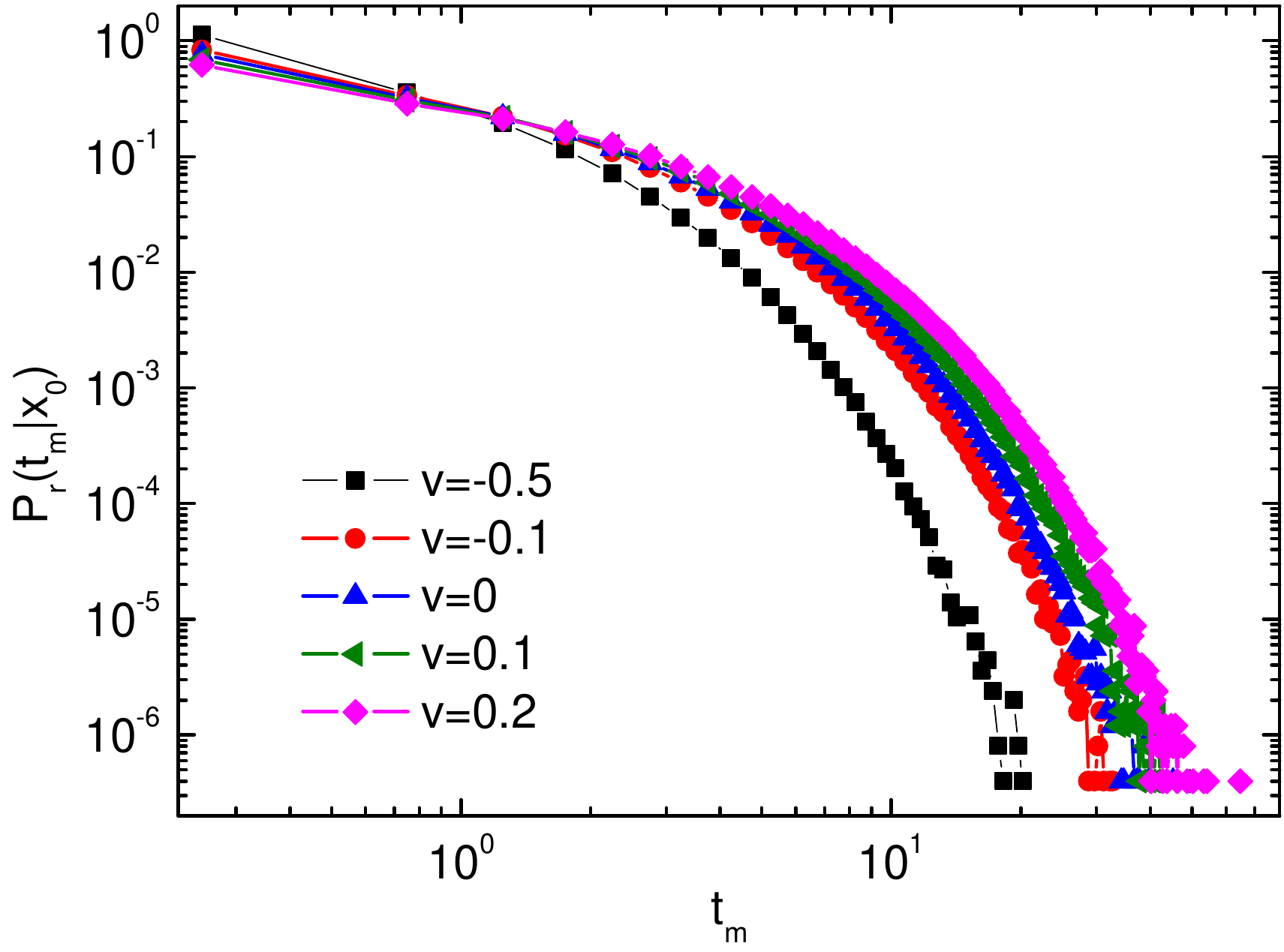}}
	\caption{\textcolor{blue}{The marginal distribution $P_r(t_m|x_0)$ of the time $t_m$ at which the maximum displacement $M$ is achieved for different drift velocity $v$, where $r=1$, $x_0=1$ and $D=1/2$ are fixed.} \label{fig5}}
\end{figure}

\textcolor{blue}{For completeness, we finally provide simulation results for the marginal distribution of $t_m$, as shown in Fig.\ref{fig5}. In the short-$t_m$ limit, $P_r(t_m|x_0)$ appears to a power-law decay, where the exponent decreases with the drift $v$. The large-$t_m$ tail of the distribution seems to be exponential. With the increase of $v$, the exponential tail decays more slowly. The qualitative behavior of $P_r(t_m|x_0)$  is similar to the previous results when the resetting is absent ($r=0$) and the drift towards the origin ($v<0$) \cite{randon2007distribution}.}

\section{Conclusions}
In conclusion, we have studied the extreme value statistics of a one-dimensional drifted Brownian motion under stochastic resetting, starting from a positive position $x_0$ and ending whenever the motion passes through the origin. We have derived the exact distribution of the maximal displacement $M$, from which we compute the mean value of $M$ as a function of the resetting rate $r$ and the drift velocity $v$. We find that both the resetting and drift have profound impacts on the distribution of $M$ and its mean value $\langle M \rangle$. For $v \geq 0$ so that the drift is away from the origin, $\langle M \rangle$ is divergent in the limit of $r \to 0$ and approaches to $2 x_0$ as $r \to \infty$. $\langle M \rangle$ is a decreasing function of $r$. While for $v<0$, the drift is towards the origin, and $\langle M \rangle$ is convergent in the limit of $r \to 0$ and also approaches to $2 x_0$ as $r \to \infty$. However, for $v<0$, $\langle M \rangle$ shows a more abundant dependence on $r$. Specifically, for $v<v_c$ where $v_c \approx -1.69415 D/x_0$, $\langle M \rangle$ is an increasing function of $r$. For $v_c<v<0$, $\langle M \rangle$ varies nonmonotonically with $r$. There is an intermediate value of $r$ at which $\langle M \rangle$ is minimized. Furthermore,  we derive in the Laplace domain the joint distribution of the maximum $M$ and the time $t_m$ at which the maximum is achieved, from which we compute the mean extreme time $\langle t_m \rangle$. Interestingly, $\langle t_m \rangle$ is a nonmonotonic function of $r$ when $v$ is larger than a critical value $v_m$, where  $v_m\approx -1.66102 D/x_0$. An optimal resetting rate occurs at which $\langle t_m \rangle$ attains its unique minimum. Otherwise, for $v<v_m$, $\langle t_m \rangle$ increases monotonically with $r$. Thus, a resetting-induced transition phenomenon is observed at $v=v_m$. Our findings suggest that the extreme time at which the maximum excursion in the drift-diffusion system can be advanced by virtue of stochastic resetting, also replenishing the phenomenon of resetting expediting random search. In the future, it would be interesting to investigate the extremal statistics of first-passage trajectories in other types of random motions under resetting, such as active Brownian motions \cite{scacchi2018mean,kumar2020active} and run-and-tumble motions \cite{evans2018run,santra2020run,bressloff2020occupation,singh2022extremal}.

\begin{acknowledgments}
This work was supported by the National Natural Science Foundation of China (11875069), the Key Scientific Research Fund of Anhui Provincial Education Department (2023AH050116), and Anhui Project (Grant No. 2022AH020009)
\end{acknowledgments}

%\bibliographystyle{apsrev}
%\bibliography{MV}

\end{document}